\documentclass[3p]{elsarticle}
\usepackage{amsmath}
\usepackage{graphicx}
\usepackage{amssymb}
\usepackage{fullpage}
\usepackage[T1]{fontenc}
\usepackage{setspace}
\usepackage{enumitem,lipsum}
\usepackage{enumitem}  
\usepackage{ragged2e}
\usepackage{mhchem}
\usepackage{wasysym}
\usepackage{booktabs}
\usepackage{threeparttable}
\usepackage{pifont}
\usepackage{natbib}
\usepackage{geometry}
\usepackage{txfonts}
\usepackage{cite}

\begin{document}
\begin{frontmatter}

\title{Magneto-rotational instability in the protolunar disk}

\author[bu]{Augusto Carballido\corref{cor1}}
\ead{Augusto{\_}Carballido@baylor.edu}

\author[sese]{Steven J. Desch}

\author[igp]{G. Jeffrey Taylor}

\cortext[cor1]{Corresponding author}

\address[bu]{Center for Astrophysics, Space Physics, and Engineering Research, Baylor University, Waco, TX 76798-7316, USA}
\address[sese]{School of Earth and Space Exploration, Arizona State University, Tempe, AZ 85287, USA}
\address[igp]{Hawai'i Institute of Geophysics and Planetology, Pacific Ocean Science and Technology (POST) Building, University of Hawai'i, 1680 East-West Road, Honolulu, HI 96822, USA}

\begin{abstract}
We perform the first study of magnetohydrodynamic processes in the protolunar disk (PLD). With the use of published data on the chemical composition of the PLD, along with existing analytical models of the disk structure, we show that the high temperatures that were prevalent in the disk would have led to ionization of Na, K, SiO, Zn and, to a lesser extent, O$_2$. For simplicity, we assume that the disk has a vapor structure. The resulting ionization fractions, together with a relatively weak magnetic field, possibly of planetary origin, would have been sufficient to trigger the magneto-rotational instability, or MRI, as demonstrated by the fact that the Elsasser criterion was met in the PLD: a magnetic field embedded in the flow would have diffused more slowly than the growth rate of the linear perturbations. We calculate the intensity of the resulting magnetohydrodynamic turbulence, as parameterized by the dimensionless ratio $\alpha$ of turbulent stresses to gas pressure, and obtain maximum values $\alpha\sim 10^{-2}$ along most of the vertical extent of the disk, and at different orbital radii. This indicates that, under these conditions, turbulent mixing within the PLD due to the MRI was likely capable of transporting isotopic and chemical species efficiently. To test these results in a conservative manner, we carry out a numerical magnetohydrodynamic simulation of a small, rectangular patch of the PLD, located at 4 Earth radii ($r_{\rm E}$) from the center of the Earth, and assuming once again that the disk is completely gaseous. We use a polytrope-like equation of state. The rectangular patch is threaded initially by a vertical magnetic field with zero net magnetic flux. This field configuration is known to produce relatively weak MRI turbulence in studies of astrophysical accretion disks. We accordingly obtain turbulence with an average intensity $\alpha\sim 7\times 10^{-6}$ over the course of 280 orbital periods (133 days at 4$r_{\rm{E}}$). Despite this relatively low value of $\alpha$, the effective turbulent diffusivity $D\sim 10^{10}-10^{11}$ cm$^{2}$ s$^{-1}$ of a passive tracer introduced in the flow is large enough to allow the tracer to spread across a radial distance of $10r_{\rm{E}}$ in $\sim13-129$ yr, less than the estimated cooling time of the PLD of $\sim250$ yr. Further improvements to our model will need to incorporate the energy balance in the disk, a complete two-phase structure, and a more realistic equation of state. 
\end{abstract}


\end{frontmatter}

\section{Introduction}

For three decades the most successful model of the Moon's origin has been the Giant Impact model 
(Benz et al., 1986), which hypothesizes that a Mars-sized impactor collided with the 
proto-Earth, ejecting mantle material into a protolunar disk from which the Moon accreted.
Through increasingly sophisticated numerical simulations, and broader parameter studies, this 
model has been able to account for the Moon's mass, its relative lack of an iron core, its 
volatile depletion, and the angular momentum of the Earth-Moon system (e.g., Canup, 2004; \'{C}uk and Stewart, 2012). 
The success of the model is such that large impacts are now accepted as a general process in 
planetary accretion, both in our solar system (e.g., Mercury: Benz et al., 1988) and in other 
planetary systems such as the 12 Myr-old system HD 172555 (Johnson et al., 2012).
Recent discoveries regarding the Moon's volatile inventories and isotopic ratios, however, have 
challenged the Giant Impact model.

\subsection{The Moon's Volatile Inventories}

The abundances of volatile elements in the Moon are particularly important for testing models of the 
Moon's origin. 
Elements can be ranked by volatility according to their 50\% condensation temperature in the solar nebula
(solar composition, $\sim 10^{-4}$ bar), as computed by Lodders (2003). 
Although the Moon formed from rock and not directly from the solar nebula, these are useful guides
for volatility in general.
Moderately volatile elements (K, Rb, and Cs) have 50\% condensation temperatures in the 800 - 1000 K range 
and have concentrations in the Moon that average $\sim 25\%$ of their abundances in the bulk silicate Earth 
(see Taylor and Wieczorek, 2014, for a recent assessment of the bulk lunar composition). 
Highly volatile elements (Bi, Zn, Cd, Br, and Tl) have 50\% condensation temperatures in the 530 - 754 K range
and are highly depleted in lunar mare basalts, KREEP basalts, and cumulate rocks from the lunar highlands 
[KREEP is an acronym for lunar rocks rich in potassium (K), rare earth elements (REE) and phosphorus (P)], 
compared to other lunar rocks.
Their average concentrations are only $\sim 1\%$ of those in the bulk silicate Earth (Taylor and Wieczorek, 2014). 
However, glassy deposits from pyroclastic eruptions are more Earth-like in their volatile abundances of highly 
volatile elements, averaging $\sim 65\%$ of those in the bulk silicate Earth (based on orange glass data from 
Morgan et al., 1974). 
The data from lunar volcanic deposits thus suggest that some regions of the lunar interior are similar to Earth 
in volatile element abundances while others are strongly depleted. 
The volatile-rich pyroclastic deposits are much less abundant on the lunar surface than are other lunar rocks, 
further suggesting that the interior is likely to be on average strongly depleted. 
However, models for lunar origin must account for the presence of some regions of non-depleted, Earth-like materials.

Besides containing Earth-like abundances of highly volatile elements, the volcanic glasses contain as much 
${\rm H}_{2}{\rm O}$ as do Mid-Ocean Ridge Basalts, or MORBs (Saal et al., 2008).
Specifically, the ${\rm H}_{2}{\rm O}$ concentrations are like those in depleted MORBs, which derived from the 
driest regions of the terrestrial mantle. 
This is significant because ${\rm H}_{2}{\rm O}$ is among the most volatile species, having a 50\% condensation 
temperature of 182 K (Lodders, 2003). 
Total ${\rm H}_{2}{\rm O}$ concentrations in other types of lunar rocks have not been quantified well, but suggest 
that concentrations inside the Moon are variable (Robinson and Taylor, 2014). 
The best estimate for the abundance of ${\rm H}_{2}{\rm O}$ in a lunar basalt is for glass in lunar basalt 15358, 
which is 5 to 10 times lower than in the magma that produced the pyroclastic glasses (Taylor and Robinson, 2015). 
Although not as well quantified as abundances of highly volatile elements, ${\rm H}_{2}{\rm O}$ concentrations seem to 
correlate roughly with abundances of the volatile elements. 
Thus, models for lunar origin need to account for the variable, possibly bimodal, abundances of volatile elements,
including water, in the Moon.

Desch and Taylor (2011) presented a simple model for the loss of volatiles from the protolunar disk (PLD). 
They demonstrated that during the short lifetime of the PLD 
($\sim 10^2$ yr; e.g., Canup, 2004), the only way to significantly deplete the PLD in any species
was through hydrodynamic escape, that is, via a thermal wind akin to the solar wind, in which the 
entire top of the disk atmosphere achieves escape velocity as a single fluid.
The rate of escape is sensitive to the Jeans escape parameter
$\lambda_{\rm{J}} = U_{\rm g} / k_{\rm B} T$, where $U_{\rm g}$ is the gravitational potential energy per gas molecule, 
$k_{\rm B}$ is Boltzmann's constant, and $T$ the gas temperature. 
Escape demands $\lambda_{\rm J} \lesssim 1$, i.e., gas that is either loosely gravitationally bound and/or 
very hot.
Desch and Taylor (2011) argued that the decrease in the gravitational potential was the more 
important effect, and that even though the gas temperature might fall off with distance from the 
Earth, escape of the disk atmosphere was much easier beyond about 5 Earth radii.
This model qualitatively suggests that material in the PLD close to the Earth 
would be relatively enriched in volatiles, and the material farther away would be depleted,
thus potentially explaining the observations of water and volatiles in the Moon.
However, this explanation demands that mixing was not extensive in the PLD.

The particular way in which the disk was formed could also have determined its final volatile distribution. Nakajima and Stevenson (2014) conducted numerical simulations of the Giant Impact and three other variant scenarios: one in which the impactor collides with a rapidly rotating Earth (\'{C}uk and Stewart, 2012), another one in which each of the colliding bodies has half an Earth's mass (Canup 2012), and one more in which the mass ratio between the bodies is 7:3. Nakajima and Stevenson (2014) found significant variation in the vapor mass fractions among the resulting disks, and although they did not take into account the rate of radial mixing, they suggest that such variation could affect the volatile content of the PLD.

\subsection{The Moon's Isotopic Abundances}

A second challenge for the Giant Impact model is to understand the nearly-but-not-quite perfect 
isotopic homogeneity in the Earth-Moon system. 
The Moon's isotopic composition is not resolvable from Earth's in many systems, including W (Touboul et al.,\ 2007),
Ti (Zhang et al.,\ 2012), Yb (Albalat et al.,\ 2012), Si (Armytage et al.,\ 2012), O (Pahlevan and Stevenson, 2007), and Cr (Lugmair and Shukolyukov, 1998). The Moon is possibly (at the $2\sigma$ level) distinct from Earth in Mg (Esat and Taylor, 1999), Fe (Poitrasson et al.,\ 2004), and K (Humayun and Clayton, 1995).
The Moon {\it is} isotopically distinct from Earth in Zn (Paniello et al.,\ 2012), N (Marty et al.,\ 2003),
and probably in H (Robinson and Taylor, 2014).
In Figure 1, these isotopic shifts between the Earth and the {\it bulk} Moon are plotted against volatility 
as measured by their 50\% condensation temperatures in the solar nebula (Lodders 2003). 
These data strongly suggest that volatile species can be isotopically fractionated, but that the 
most refractory elements are not. 

\begin{figure}
\centering 
\noindent\includegraphics[width=0.65\textwidth]{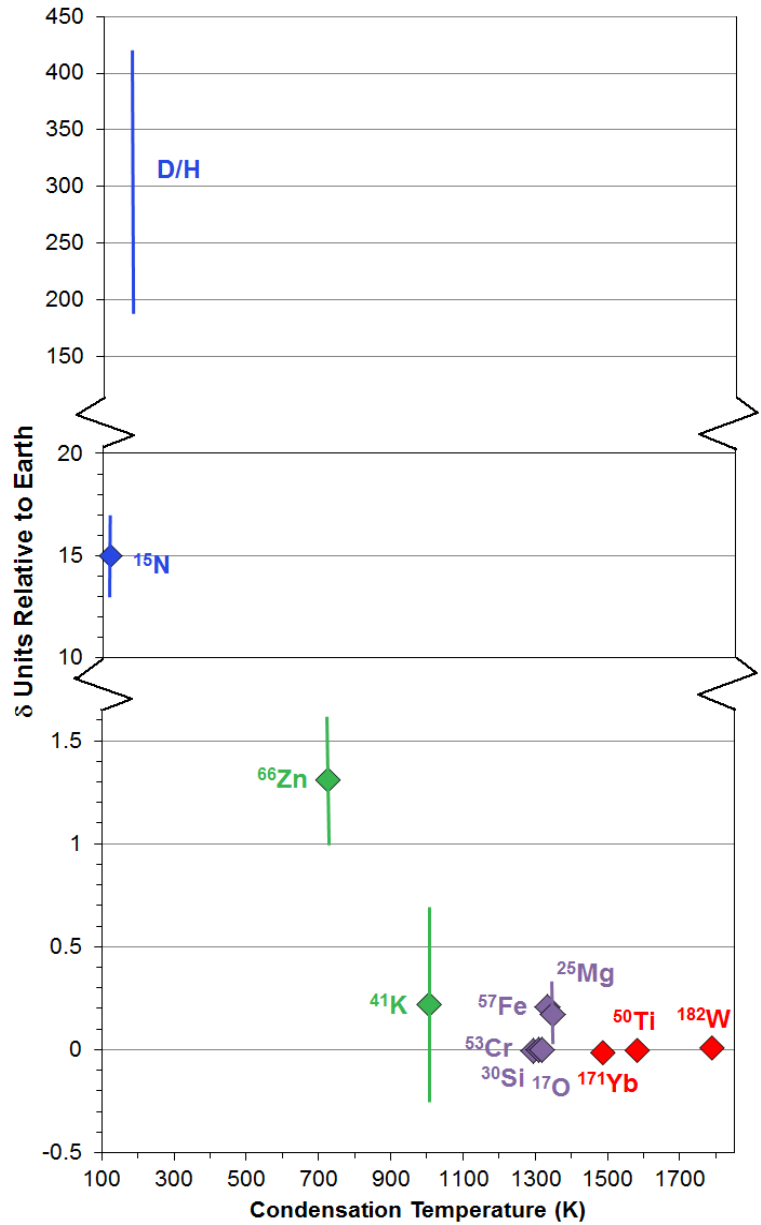} 
\caption{Deviation of isotopic compositions (in parts per thousand) of elements in the bulk Moon,
compared to Earth (a perfect match would be 0 per mil).  Error bars show the $2\sigma$ variation
from average values, except for D/H, which shows the range in samples with minimal outgassing.
Uncertainties for elements without error bars are about the size of the symbols.  Isotopic deviations
are plotted against volatility, measured using the 50\% condensation temperature in a low-pressure 
solar-composition gas (O is assumed to condense with major elements).  Strongly refractory elements 
(red) and major elements (purple) deviate only slightly from terrestrial values.  Moderate volatiles 
(green) are distinctly fractionated (Zn, enriched in heavy isotopes) or possibly not fractionated (K). 
The volatile element N appears fractionated, and highly volatile hydrogen is distinctly enriched in 
the heavy isotope, although by how much is uncertain.  Data sources in the text. }
\label{fig:isotopes}
\end{figure}

Isotopic fractionation during volatile loss is understandable; e.g., it might be fractionated during 
outgassing of the volatile from magma, followed by subsequent loss of the outgassed volatile in a
hydrodynamic wind, as described above. 
But the remarkable isotopic similarity between the Moon and Earth in the more refractory elements is 
difficult to explain. 
There are three possible explanations: either the Earth and the impactor formed from isotopically similar materials;
or they were isotopically distinct but contributed equally to the PLD;
or they were different and contributed unequally, but the PLD mixed with the Earth during its
lifetime. 
The first explanation is advocated by Mastrobuono-Battisti et al.\ (2015), who point out that the Earth 
and the impactor probably formed at similar annuli in the solar nebula (but see Kaib and Cowan, 2015 for an opposing view). 
The second explanation is pursued in new simulations of the collision by \'{C}uk and Stewart (2012).
The last explanation was championed by Pahlevan and Stevenson (2007), who argued that turbulent mixing could
potentially mix material throughout the disk, and possibly between the Earth and the disk. 

If the PLD is found to be well mixed, that would support this explanation, and would
constrain disk processes at play as the Moon formed.
If it is found to be not well mixed, then one of the first two explanations would be supported, and
constraints could be placed on the isotopic abundances of the impactor, or on the nature of the collision
itself. 
Arguments made by Desch and Taylor (2013) suggest that mixing between the Earth and the PLD
would be vigorous, at least initially (until a laminar angular velocity profile between the Earth and
the disk is set up), but the rate of mixing within the PLD itself is not known.

\subsection{A new mixing mechanism?}

The Moon's inventories of volatile elements, and its isotopic abundances, all relate to or constrain the degree of mixing in the PLD. If the Earth-Moon isotopic homogeneity arose by mixing, then the PLD was
exceptionally well mixed on short timescales. 
Determining the mixing rate in the PLD has profound implications for the provenance of the impactor, 
for the dynamics of the impact, and the origin of the Moon generally.

Previous studies of mixing in the PLD (e.g., Pahlevan and Stevenson 2007) have considered turbulent mixing in which
 the effective viscosity was parameterized using an assumed value of $\alpha$, the dimensionless turbulence parameter of
 Shakura and Sunyaev (1973).  An improved approach would be to estimate the mixing rate from first principles, considering the physical
 state of the disk.
 
One important aspect of the PLD that bears on mixing, that has not been considered in previous work, is the ionization state of the disk. Analytical models and numerical simulations of the lunar-forming Giant Impact produce disks whose temperatures are above $\sim$2000 K (Thompson and Stevenson, 1988, hereafter TS88; Canup, 2004; \'{C}uk and Stewart, 2012). At these temperatures, thermal ionization could give rise to a significant electron fraction in the gaseous component of the PLD. Unlike protoplanetary (Umebayashi and Nakano, 1988; Balbus and Hawley, 2000; Fromang et al., 2002; Turner et al., 2014a) and giant-planet circumplanetary (Fujii et al., 2014; Keith and Wardle, 2014; Turner et al., 2014b) disks, for which thermal ionization is generally confined to the vicinity of the central object, the thermal structure of the PLD may have rendered most of the vapor phase of the disk ionized.

This is important for the following reason: any astrophysical disk whose angular velocity decreases with distance from the center, regardless of how it formed, may be subject to the fluid instability known as \textit{magneto-rotational instability}, or MRI, if sufficient ionization and a weak magnetic field are present (Balbus and Hawley, 1991a,b; Balbus and Hawley, 1998). In essence, if a magnetic field line is anchored to two parcels of ionized disk gas on different orbits, the line will be stretched by the differential rotation of the system, and magnetic tension will transfer angular momentum from the inner gas element to the outer one. During this process, the disk fluid becomes turbulent, and transport of material can occur along the radial direction of the disk. It is important to note that the MRI is a \textit{local} mechanism within the disk.

In this work, we evaluate the feasibility that the PLD was susceptible to the MRI. Using two previous analytical models of the PLD as templates (TS88; Ward, 2012), a standard criterion for MRI activity is borrowed from analyses of protoplanetary and circumplanetary disks, in an effort to determine the likelihood of PLD magnetohydrodynamic (MHD) turbulence. Note that we do not address processes that may arise from the interaction between the PLD and the post-impact Earth. 

It is not generally assumed that the PLD  was threaded by a magnetic field, although we consider it plausible.  If either the proto-Earth
 or its impactor, Theia, possessed an internal field, we view it as  likely that they would have retained that magnetization during the impact, exposing
 the PLD to magnetic fields. It is unclear whether the Earth possessed a geodynamo at the time of the Giant Impact. Models predict one
 due to the presence of a primitive, electrically conducting basal magma ocean surrounding Earth's core (Ziegler et al., 2013), but the
 oldest available evidence comes from 3.45 billion-year-old quartz phenocrysts (Aubert et al., 2010), 1.02 Gy younger than the calculated
 age of the Moon-forming impact at 4.47 Ga (Jacobsen et al., 2014). Another, largely unexplored possibility, is that the Giant Impact
 itself may have generated magnetic fields (e.g., Crawford and Schultz, 1999).

In the following, we assume that a circumterrestrial disk has been produced as a result of the Giant Impact, irrespective of the exact collision parameters. Furthermore, we assume that the disk is partially threaded by magnetic field lines. 

The outline of this paper is as follows. In Sec. \ref{sec:model} we present the relevant results of existing analytical models of the PLD that we will use for our calculations. In Sec. \ref{sec:ion} we perform an assessment of the ionization conditions in a particular instance of the PLD. We show in Sec. \ref{sec:turb} that those conditions are conducive to turbulence generated by the MRI, and we compute the intensity of such turbulence. Section \ref{sec:MHDsim} presents the results of a numerical MHD simulation of a local patch of the PLD; it shows that turbulent diffusivity may have operated on short enough time scales to allow for isotopic homogenization in the PLD. In Sec. \ref{sec:discussion} we comment on different considerations and improvements for future work, and conclusions are offered in Sec. \ref{sec:conc}.

\section{Protolunar disk model}\label{sec:model}
TS88 derived a detailed analytical model of the structure of the PLD. This model describes the disk as a two-phase medium, vapor and liquid, which is the product of the ejecta from the Giant Impact. The liquid component, the bulk of which resides close to the disk midplane, is assumed to be SiO$_2$, while the vapor component is a mixture of SiO and O$_2$. Their PLD model hovers on the brink of gravitational instability, as a result of radiative photospheric losses of the energy dissipated within the disk. The metastable state allows the disk to spread beyond the Roche boundary without fragmenting, and its surface density decreases in the process. The onset of fragmentation into possible Moon-forming blobs occurs when the surface density decreases so much that gravitational instability is needed for viscous dissipation. 

By examining the vertical structure of the PLD, Ward (2012) suggested that the critical surface density can be reached through condensation of magma droplets from the vapor phase. This would compensate energy losses by the release of latent heat. In this manner, the disk vapor component would be depleted in $\sim$ 250 years.

As a first step in our analysis of magnetic effects in the protolunar disk, we need to determine its temperature profile, in order to assess the degree of ionization. The thermal structure of the PLD was calculated by TS88, who employed the phase equilibrium Clausius-Clapeyron relation

\begin{equation}\label{eq:claus}
P = P_0 \mathrm{e}^{-T_{0}/T}
\end{equation}

\noindent to approximate the response of the system pressure $P$ to the temperature $T$, with $P_0=3\times10^{14}$ dyne cm$^{-2}$ and $T_0 = 6\times 10^{4}$ K. They further estimated the photospheric temperature as a function of distance $r$ from the center of the Earth as

\begin{equation}\label{eq:Tphot}
\begin{split}
T_{\textrm{phot}} & = T_0 \left[\ln \left(\frac{P_0}{P_{\textrm{phot}}}\right)\right]^{-1} \\
& \approx 2200\textrm{K} \left\{1 + 0.036\ln\left( \frac{\kappa}{\textrm{1 cm$^2$ g$^{-1}$} } \right) + 0.036 \ln \left[\frac{H_{\textrm{phot}}}{r_{\textrm{E}}} \left(\frac{r}{r_{\textrm{E}}} \right)^3 \right] \right\}^{-1} 
\end{split}
\end{equation}

\noindent where $P_{\textrm{phot}}$ is the photospheric pressure, $\kappa$ is the photospheric opacity (for which an uncertain nominal value of 1 cm$^2$ g$^{-1}$ is adopted), $H_{\textrm{phot}}$ is the height of the photosphere, and $r_{\textrm{E}}$ is Earth's radius. The photospheric height in Eq. (\ref{eq:Tphot}) is given by the ratio of the disk sound speed to its angular velocity, 

\begin{equation}\label{eq:Hphot}
H_{\textrm{phot}}\sim \frac{c_{\textrm{s}}}{\Omega} 
\end{equation}

\noindent Now, TS88 use the following expression for the sound speed derived for the case of a two-phase medium:

\begin{equation}\label{eq:cs}
c_s = \frac{x + \left(1-x\right)\left(\rho/\rho_l\right) }{\left[ \left(\mu/RT - 2/\ell + C_g T/\ell^2 \right)x + \left(C_l T/\ell^2\right) \left(1 - x \right) \right]^{1/2}},
\end{equation}

\noindent where $x$ is the fraction of the total disk mass in the gas phase, $\rho$ and $\rho_l$ are the gas and liquid density, respectively, $\mu\approx 30$ g mol$^{-1}$ is the mean molecular weight of the gas, $R$ is the universal gas constant, $\ell=1.7\times 10^{11}$ erg g$^{-1}$ is the silicate latent heat of vaporization, and $C_g$ and $C_l$ are the specific heats of the gas and liquid phases. In the present treatment, we wish to examine possible MHD effects \textit{only} in the gas component of the PLD, and therefore we direct our attention to the limit $x\rightarrow 1$. In this case, the sound speed $c_s \rightarrow \sqrt{RT/\mu}$, and Eq. (\ref{eq:Hphot}) becomes

\begin{equation}\label{eq:Hphot2}
H_{\textrm{phot}} \sim \Omega^{-1} \left(\frac{RT_{\textrm{phot}}}{\mu}\right)^{1/2}.
\end{equation}

From Eq. (\ref{eq:claus}), the photospheric temperature can be written as $T_{\textrm{phot}}=T_0\left[ \ln\left(P_0/P_{\textrm{phot}}\right)\right]^{-1}$, with $P_{\textrm{phot}}=\Omega^2 H_{\textrm{phot}}/\kappa$. Substituting in Eq. (\ref{eq:Hphot2}) one obtains

\begin{equation}\label{eq:Hphot3}
H^2_{\textrm{phot}}=\frac{RT_0}{\Omega^2 \mu}\left[\ln\left(\frac{P_0 \kappa}{\Omega^2 H_{\textrm{phot}}}\right) \right]^{-1}.
\end{equation}

\noindent Notice that, as pointed out by TS88, disregarding the logarithm yields a photospheric scale height that increases with radius as $r^{3/2}$. 

The numerical solution to Eq. (\ref{eq:Hphot3}) can then be used in Eq. (\ref{eq:Tphot}) to obtain the radial dependence of the photospheric temperature (Fig. 4 of TS88). 

Ward (2012) obtained an approximation to the PLD temperature as a function of height $z$, given by

\begin{equation}\label{eq:Tz}
T \approx T_{\textrm{c}}\left[1 - \frac{1}{x_{\textrm{c}}}\left(\frac{z}{H_0}\right)^2\right].
\end{equation}
 
\noindent In this expression, $T_{\textrm{c}}$ is the mid-plane temperature, $H_0\equiv \left(2\ell\right)^{1/2}/\Omega$, and $ x_{\textrm{c}}$ is the gas mass fraction at the disk mid-plane. The disk profiles formulated by Ward (2012) assumed three different values of $x_{\textrm{c}}$, with their corresponding values of $T_{\textrm{c}}$. Here we will use $x_{\textrm{c}}=1$ and $T_{\textrm{c}}=4218$ K. This temperature profile is shown in Fig. 5 of Ward (2012).

\section{Ionization fractions in the PLD}\label{sec:ion}
 Temperatures in the PLD have been calculated to be at least 2000 K (TS88, Ward, 2012) and as high as $\sim10,000$ K (\'{C}uk and Stewart, 2012). These values are suggestive of thermal ionization in the disk gas. In order to evaluate the feasibility of this process, we use the Saha equation,
 
 \begin{equation}\label{eq:saha1}
 \frac{n_{\rm{i}}n_{\rm{e}}}{n} = \frac{2 g_{\rm{i}}}{g}\left(\frac{2\pi m_{\rm{e}} k_{\rm{B}} T}{h^2}\right)^{3/2} \exp{\left(-\frac{\chi}{k_{\rm{B}}T}\right) },
 \end{equation}

\noindent where $n_{\rm{i}}$ and $n$ are, respectively, the number densities of the ionized and neutral components of the species being considered, and $n_{\rm{e}}$ ($=n_{\rm{i}}$) is the number density of electrons. The partition functions, of order unity, for the ions and neutrals are $g_{\rm{i}}$ and $g$. The electron mass is $m_{\rm{e}}$; $k_{\rm{B}}$ is the Boltzmann constant; and $h$ is Planck's constant. Lastly, $\chi$ is the ionization potential of the species. 

Dividing Eq. (\ref{eq:saha1}) through by $n_{\rm{n}}$, the number density of all neutral species present, we have

\begin{equation}\label{eq:saha2}
\frac{n_{\rm{i}}}{n} \frac{n_{\rm{e}}}{n_{\rm{n}}} \approx \frac{2.41 \times 10^{15}}{n_{\rm{n}}} T^{3/2} \exp{\left(-\frac{\chi}{k_{\rm{B}}T}\right)},
\end{equation}

\noindent and defining $\xi \equiv \frac{n_{\rm{e}}}{n_{\rm{n}}}$ and $f\equiv \frac{n}{n_{\rm{n}}}$, we can re-write Eq. (\ref{eq:saha2}) as  

\begin{equation}\label{eq:saha3}
\xi^{2} \approx f \frac{ 2.41 \times 10^{15}}{n_{\rm{n}}} T^{3/2} \exp{\left(-\frac{\chi}{k_{\rm{B}}T}\right)}
\end{equation}

\noindent where we have used the charge neutrality condition $n_{\rm{e}} = n_{\rm{i}}$.

\begin{figure}[h!]
\begin{center}
\includegraphics[width=0.9\textwidth]{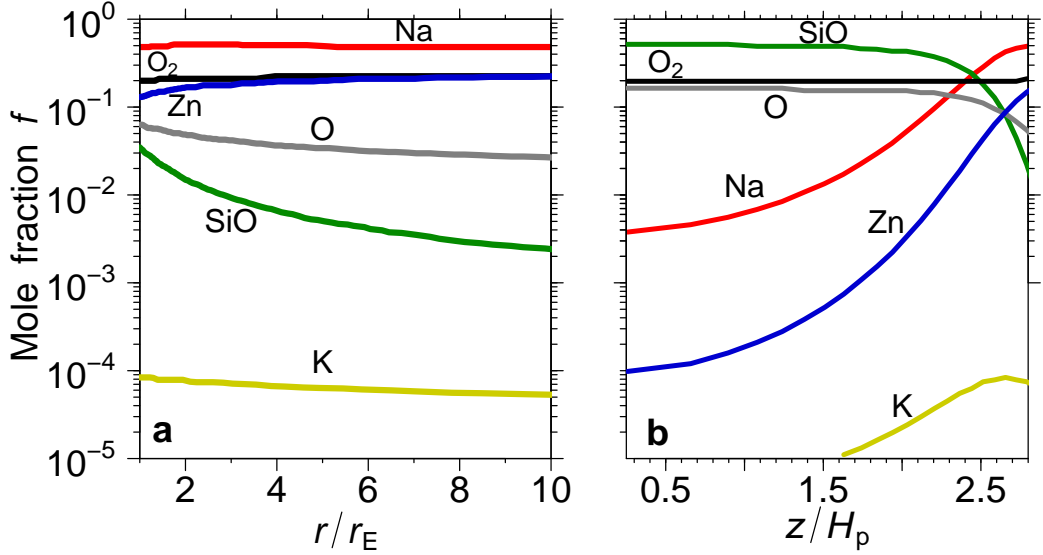}
\end{center}
\caption{\small{\textbf{(a) }Radial and \textbf{(b)} vertical profiles of chemical abundances in the protolunar disk, using data by Visscher and Fegley (2013) and the disk models of TS88 and Ward (2012). Abundances for five species are shown. Disk radial distances are given in units of Earth's radius $r_{\rm{E}}$, and vertical distances are given in units of the disk scale-height $H_{\rm{p}}$. The values shown in the left panel correspond to a disk height near the photosphere, and those on the right panel correspond to an orbital radius $r=2r_{\rm{E}}$.}    \label{fig:molefracs}}
\end{figure}

Next, we need to determine what chemical species to take into account. Visscher and Fegley (2013) computed the chemical composition of the PLD through thermochemical equilibrium calculations of a silicate melt and coexisting vapor. They obtained abundances in the saturated silicate disk atmosphere as a function of temperature, between $\sim1800$ and 4200 K (their Fig. 2). We would like to express those abundances as a function of distance from both the Earth and the disk mid-plane. In this way, we will obtain values of $f$ to substitute into Eq. (\ref{eq:saha3}). To that end, we solve Eq. (\ref{eq:Tphot}) for the radial distance $r$, and Eq. (\ref{eq:Tz}) for the vertical distance $z$, both in terms of the temperature. By digitizing the data for the chemical abundance curves from Visscher and Fegley (2013), corresponding to the bulk silicate Moon, we obtain the PLD radial and vertical profiles of mole fractions $f$ shown in Figs. \ref{fig:molefracs}a and \ref{fig:molefracs}b, respectively. The data in Fig. \ref{fig:molefracs}a correspond to a photospheric height in the disk, while the values in Fig. \ref{fig:molefracs}b correspond to an orbital radius $r=2r_{\rm{E}}$.

One thing to note is the prominent role of Na (red curves). Despite its relatively low abundance along most of the disk's vertical extent (except in the uppermost regions, Fig. \ref{fig:molefracs}b), this element is essential in establishing the ionization fraction of the disk due to its low ionization potential of 5.14 eV. Something similar occurs with K (yellow curves), which has an ionization potential of 4.34 eV  [we have also considered the effects on the ionization state due to thermionic and ion emission from the magma (Desch and Turner, 2015); we find that at temperatures above 1500 K, gas-phase thermal ionizations and recombinations of alkalis,
  especially K, dominate]. 

We are now in a position to calculate the ionization fraction $\xi$ in the PLD. We will do so for the six chemical species of Fig. \ref{fig:molefracs}. The total neutral number density $n_{\rm{n}}$ in Eq. (\ref{eq:saha3}) can be expressed as $n_{\rm{n}} = \rho/m$, where the gas density $\rho$ depends either on $r$ or on $z$, and $m$ is the mass of one atom or molecule. 

For the radial dependence of the gas density, we employ an expression derived by TS88 for the disk surface density:

\begin{equation}\label{eq:sigmaf}
\Sigma(r) \approx 1.5\times 10^{6}\left(\frac{r}{r_{\rm{E}}}\right)^{-1/2} \left\{1 + 0.036 \ln\left[ \frac{H_{\rm{phot}}}{r_{\rm{E}}} \left( \frac{r}{r_{\rm{E}}} \right)^{-3} \right] \right\}^{-4/3} \, \rm{g} \, \rm{cm}^{-2} .
\end{equation}

\noindent In a strict sense, this value of the PLD surface density corresponds to that at which gravitational instability grows in an orbital period. The disk density then becomes $\rho = \Sigma/\left(\sqrt{2\pi} H_{\rm{phot}}\right)$, and 

\begin{equation}\label{eq:nnr}
n_{\rm{n}}(r) = \frac{\Sigma(r)}{\sqrt{2\pi}m H_{\rm{phot}}(r)}.
\end{equation}

The vertical dependence of the gas density is obtained by combining the ideal gas equation of state, $P=\rho R T/\mu$, and the Clausius-Clapeyron equation (\ref{eq:claus}):

\begin{equation}\label{eq:rhoClaus}
\rho(z) = \left(\frac{P_0 T_0}{\ell}\right)\frac{1}{T(z)}\mathrm{e}^{-T_0/T(z)}, 
\end{equation}

\noindent where $T(z)$ is given by Eq. (\ref{eq:Tz}). Then

\begin{equation}\label{eq:nnz}
n_{\rm{n}}(z) = \frac{\rho(z)}{m}. 
\end{equation}

\noindent Equations (\ref{eq:nnr}) and (\ref{eq:nnz}) can now be used in the ionization equation (\ref{eq:saha3}), in which the temperature $T$ will be given by Eq. (\ref{eq:Tphot}) or (\ref{eq:Tz}), respectively.

Finally, we need to determine whether the values given by Eq. (\ref{eq:saha3}) will be high enough for the disk gas to couple to a magnetic field, and hence for the onset of the MRI. Two criteria need to be met. The first criterion requires that the instability grows faster than the magnetic field can diffuse across the wavelength of the instability's fastest-growing mode. This can be written as 

\begin{equation}\label{eq:elsasser}
\Lambda \equiv \frac{v_{\textrm{A},z}^2}{\eta \Omega} > 1,
\end{equation}

\noindent where $\Lambda$ is the Elsasser number; $v_{\rm{A},z}=B_z/\sqrt{4\pi \rho}$ is the Alfv\'{e}n speed of the vertical component of the magnetic field, $B_z$; and $\eta$ is the magnetic diffusivity. Strictly speaking, Eq. (\ref{eq:elsasser}) is the Elsasser criterion due to Ohmic dissipation, and it is found to provide a better estimation for the appearance of turbulence than, for example, the magnetic Reynolds number $Re_{\textrm{m}}\equiv c_{s}^2/\eta \Omega$, because a critical value for the latter must be determined for each magnetic field configuration, whereas the Elsasser number explicitly shows the magnetic field dependence (Turner et al., 2007). For a discussion of the possible role of ambipolar diffusion, see Section \ref{sec:discussion}.

The second criterion is that the wavelength of the fastest-growing MRI mode be smaller than the disk scale height: 

\begin{equation}\label{eq:lambda}
\lambda < H_{\rm{p}}.
\end{equation}

\noindent Note that this criterion may be slightly modified if the magnetic field itself is enhanced by the MRI, and additional considerations may have to be taken when accounting for the vertical variation of the Alfv\'{e}n speed and other quantities (Okuzumi and Ormel, 2013). 

We would like to ascertain whether condition (\ref{eq:elsasser}) holds for the chemical species that we considered above. To this end, we need to calculate the magnetic diffusivity in the PLD. The magnetic diffusivity is inversely proportional to the electrical conductivity $\sigma_{\rm{c}}$:

\begin{equation}\label{eq:eta}
\eta = \frac{c^2}{4\pi \sigma_{\rm{c}}}.
\end{equation}

\noindent In this expression, $c$ is the speed of light, and 

\begin{equation}\label{eq:elec_cond}
\sigma_{\rm{c}} = \frac{e^2 n_{\rm{e}}}{m_{\rm{e}} \nu_{\rm{e}}},
\end{equation}

\noindent with $e$ the elementary charge and $\nu_{\rm{e}}$ the electron-neutral collision frequency. If $\sigma$ is the collision cross-section, then

\begin{equation}\label{eq:colfreq}
\nu_{\rm{e}} = n_{\rm{n}}\langle \sigma v_{\rm{ne}} \rangle_{\rm{e}}.
\end{equation}

\begin{figure}[h!]
\begin{center}
\includegraphics[width=0.9\textwidth]{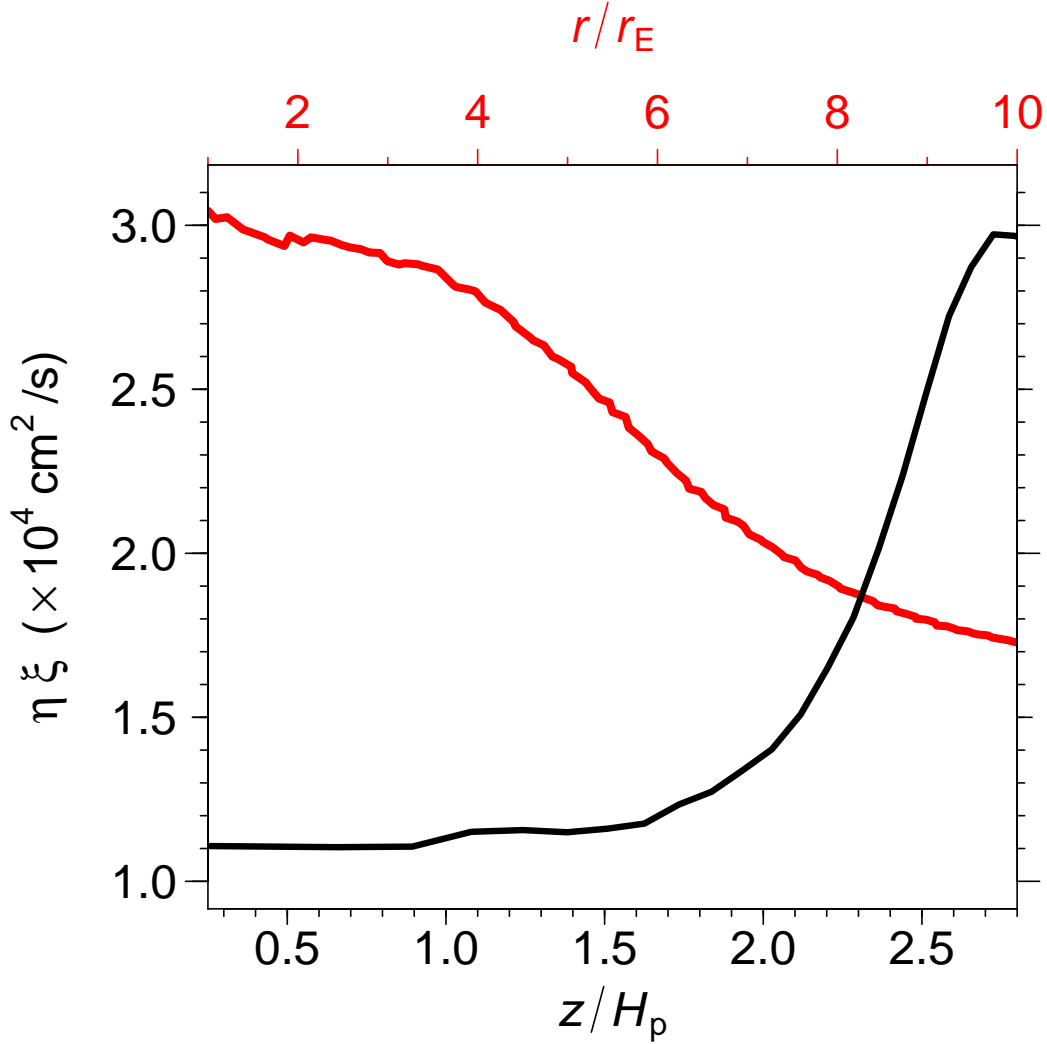}
\end{center}
\caption{\small{Radial (\textit{red} curve, top horizontal axis) and vertical (\textit{black} curve, bottom horizontal axis) profiles of the product between magnetic diffusivity $\eta$ and ionization fraction $\xi$ in the protolunar disk, calculated from Eqs. (\ref{eq:eta})-(\ref{eq:cross_sec}). Expressions (\ref{eq:Tphot}) and (\ref{eq:Tz}) were used for the temperature $T$. The radial profile corresponds to the height of the photosphere, and the vertical profile corresponds to an orbital radius $r=2r_{\rm{E}}$ }}    \label{fig:etaxi}
\end{figure}

\noindent The brackets in the collision rate $\langle \sigma v_{\rm{ne}} \rangle_{\rm{e}}$ denote a velocity-weighted average. Numerical values for the collision rate in a H/H$_2$/He gas have been calculated previously (Draine et al. 1983), and are commonly used for protoplanetary disks. Of course, the composition of the PLD requires that we re-calculate those values. For each chemical species $j$, we approximate the collision cross section as 

\begin{equation}\label{eq:cross_sec}
\sigma \approx \frac{\pi \sum\limits_{j} f_j\left(r_j+r_{\rm{e}}\right)^2}{\sum\limits_{j} f_j}, 
\end{equation}

\noindent where $r_j$ is either the van der Waals radius of the atom in question or the bond length of the molecule being considered, $r_{\rm{e}}$ is the classical electron radius, and $f_j$ the mole fraction of species $j$, calculated by Visscher and Fegley (2013). We include iron and magnesium atoms in this estimate of $\sigma$, since they are relatively abundant in the Visscher and Fegley model. Using $v_{\rm{ne}}=\left(8\pi k_{\rm{B}} T/\mu_{j\rm{e}}\right)^{1/2}$, where $\mu_{j\rm{e}}$ is the reduced mass of the atom or molecule and the electron ($\mu_{j\rm{e}}\approx m_{\rm{e}}$), and with $T$ given by Eq. (\ref{eq:Tphot}) or (\ref{eq:Tz}), we can then obtain the magnetic diffusivity from Eqs. (\ref{eq:eta})-(\ref{eq:cross_sec}), or, more precisely, the product $\eta \xi$. The radial and vertical profiles of this product are plotted in Fig. \ref{fig:etaxi}.

To obtain a value of $v_{\textrm{A},z}$ for Eq. (\ref{eq:elsasser}), we use condition (\ref{eq:lambda}). The fastest-growing MRI wavelength is

\begin{equation}\label{eq: lambdava}
\lambda_{\rm{MRI}}\approx \frac{2\pi v_{\textrm{A},z}}{\Omega}.
\end{equation}

\noindent The disk pressure scale height is given by Ward (2012) as

\begin{equation}\label{eq:hp}
H_{\rm{p}}=\left(\frac{2RT_{\rm{c}}x_{\rm{c}}}{\mu}\right)^{1/2} \Omega^{-1}.
\end{equation}

\noindent Recall that $x_{\textrm{c}}=1$ and $T_{\textrm{c}}=4218$ K. Condition (\ref{eq:lambda}) then gives

\begin{equation}\label{eq:vamax}
v_{\textrm{A},z} \lesssim 2.43\times 10^{4} \, \rm{cm} \, \rm{s}^{-1}.
\end{equation}

\noindent Combining Eqs. (\ref{eq:elsasser})-(\ref{eq:vamax}), we obtain the critical value of the ionization fraction above which the MRI can develop:

\begin{equation}\label{eq:xicrit}
\xi_{\textrm{crit}} \approx \left(1.49 \times 10^{12} \, \rm{cm}^{-2}\right) \left(\frac{\Omega}{4.38\times 10^{-4} \, \rm{s}^{-1}}\right) \left(\frac{T}{2000\,\rm{K}}\right)^{1/2} \sigma.
\end{equation}

\begin{figure}[h!]
\begin{center}
\includegraphics[width=0.9\textwidth]{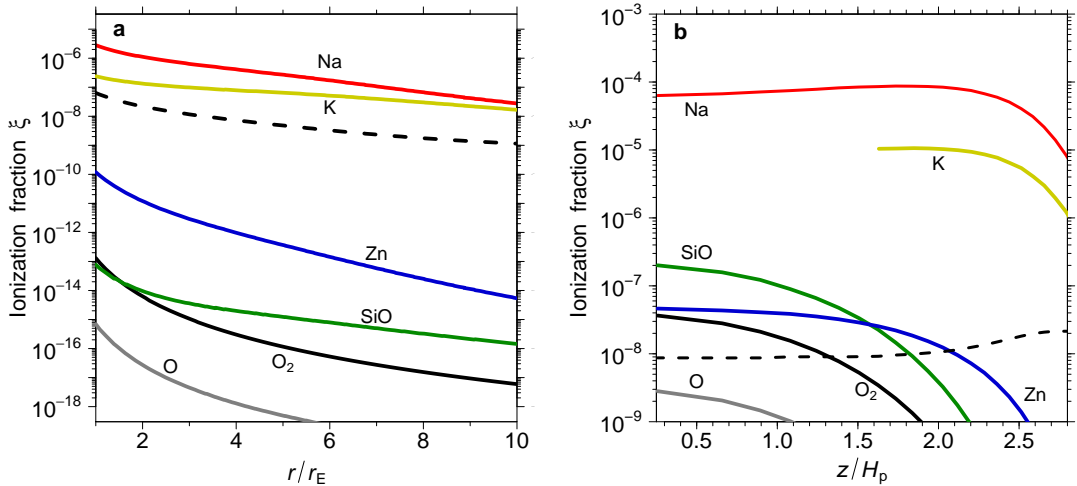}
\end{center}
\caption{\small{\textbf{(a) }Radial and \textbf{(b)} vertical profiles of ionization fraction (solid curves) in the PLD for the chemical species of Fig. \ref{fig:molefracs}.  The dashed curves are the critical values, given by Eq. (\ref{eq:xicrit}), above which the gas is sufficiently ionized for magnetic coupling, and hence for the development of the MRI. As in Fig. \ref{fig:molefracs}, the values shown in the left panel correspond to a disk height near the photosphere, and those on the right panel correspond to an orbital radius $r=2r_{\rm{E}}$. }    
\label{fig:ionfracs}}
\end{figure}

The profiles of ionization fraction derived from the above calculations [Eqs. (\ref{eq:saha3}) and (\ref{eq:xicrit})] are shown in Fig. \ref{fig:ionfracs}. Figure \ref{fig:ionfracs}a shows the radial profiles of ionization fraction near the PLD photosphere, and Fig. \ref{fig:ionfracs}b shows the vertical profiles at $r=2r_{\rm{E}}$. The solid curves represent the values of $\xi$, while the dashed curves stand for the critical values $\xi_{\rm{crit}}$. As can be seen in Fig. \ref{fig:ionfracs}a, sodium and potassium are the main contributors to thermal ionization in the upper layers of the PLD, all along the radial extent of the disk. Figure \ref{fig:ionfracs}b also shows the prominent contribution of these two elements to the ionization of a disk column [the cutoff for the potassium curve at about $z=1.5H_{\rm{p}}$ is due to unavailable data above 3500 K in the Visscher and Fegley (2013) calculations for a bulk silicate Moon composition]. Thermal ionization of Na and K is so efficient that the PLD behaves as an ideal MHD system. Figure \ref{fig:ionfracs}b also reveals that Zn, SiO, and to a lesser extent, O$_2$, keep the PLD sufficiently ionized for magnetic coupling in the vicinity of the midplane.

\section{Turbulence in the PLD}\label{sec:turb}
The viscous evolution of an accretion disk is parameterized by a dimensionless number, conventionally called $\alpha$. This  number provides a link between the viscosity generated by turbulent fluid motions in the disk, $\nu_{\rm{t}}$, and the disk structure, represented by the disk scale height $H$ and the disk sound speed $c_{\rm{s}}$:

\begin{equation} \label{eq:alphat}
\nu_{\rm{t}}=\alpha c_{\rm{s}} H.
\end{equation} 

\noindent This equation expresses the fact that the outermost scale of the turbulent flow is bounded by the smallest scale in the disk, $H$, and the velocity of the turbulent motions is limited to be no larger than the sound speed, since supersonic motions would result in shocks and rapid dissipation. In effect, the turbulent parameter $\alpha$ measures the efficiency of angular momentum transport in accretion disks, and the prescription given by Eq. (\ref{eq:alphat}) restricts $\alpha$ to be less than unity. It is important to note that $\alpha$ is not necessarily a constant (in fact, it is unlikely to be, as illustrated in Section \ref{sec:MHDsim} below). 

Pahlevan and Stevenson (2007) argued that convective turbulence would have homogenized the isotopic composition of the PLD, if turbulent mixing was able to operate efficiently, and obtained an equilibration time scale of $\sim10^{2}$--$10^{3}$  yr for $\alpha \sim 10^{-4}$--$10^{-3}$. From the models of TS88 and Ward (2012), and considerations of the development of the MRI, we can also estimate values of $\alpha$ that we can compare to those calculated for convection.

\begin{figure}[h!]
\begin{center}
\includegraphics[width=0.99\textwidth]{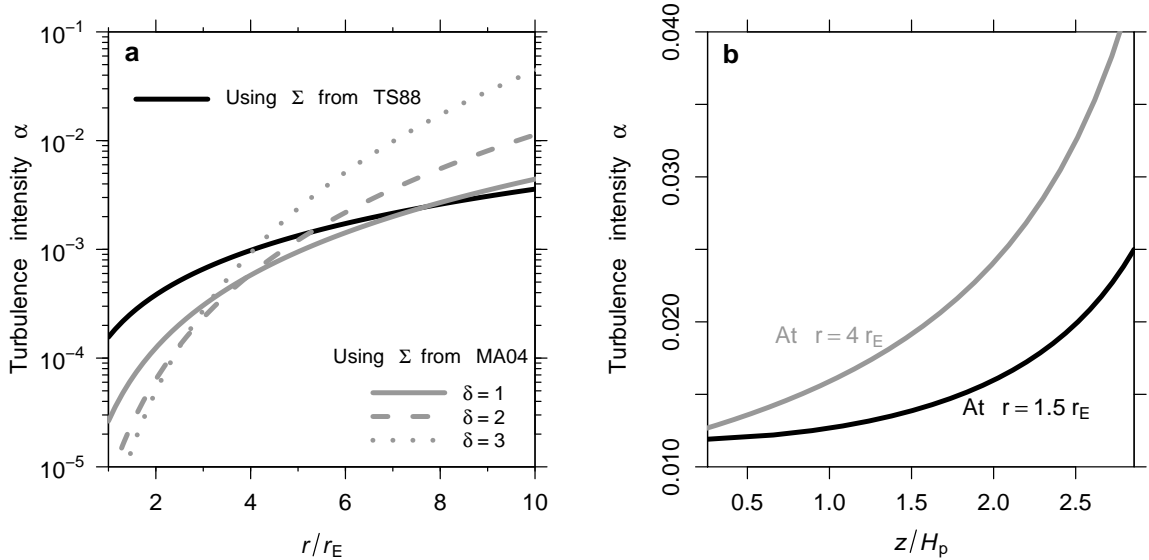}
\end{center}
\caption{\small{\textbf{(a)} Radial profiles of the turbulent intensity $\alpha$ in the PLD derived by balancing viscous dissipation against photospheric losses, and calculated from Eq. (\ref{eq:alpha}) along with two different expressions for the disk surface density: Eq. (\ref{eq:sigmaf}) by TS88 (\textit{solid, black curve}), and Eq. (\ref{eq:sigmama1}) by MA04 (\textit{gray curves}); \textbf{(b)} Maximum, plausible vertical profiles of $\alpha$ due to the MRI, calculated at $r=1.5r_{\rm{E}}$ (\textit{black curve}) and at $r=4r_{\rm{E}}$ (\textit{gray curve}), from upper limits in inequality (\ref{eq:az}).  }    \label{fig:alpha}}
\end{figure}

By assuming that the viscous dissipation in the disk exactly matches the photospheric losses [$(9/4) \Sigma \nu_{\rm t} \Omega^2 = \sigma_{\rm{SB}}  T_{\rm{phot}}^4$, with $\sigma_{\rm{SB}}$ the Stefan-Boltzmann constant], TS88 obtained the radial profile of $\alpha$ as

\begin{equation}\label{eq:alpha}
\alpha(r) = 6\times 10^{-5}\left(\frac{r}{3r_{\rm{E}}}\right)^{3/2}\left[\frac{\Sigma(r)}{10^{7}\textrm{~g cm$^{-2}$}}\right]^{-1}\left[\frac{T_{\rm{phot}}(r)}{2000\textrm{~K}}\right]^{4}\left[\frac{T_{\rm{mid}}(r)}{3000\textrm{~K}}\right]^{-1},
\end{equation}

\noindent where $T_{\rm{mid}}(r)=T_0/\ln(P_0/P_{\rm{mid}})$ is the mid plane temperature, with $P_{\rm{mid}}$ the mid plane pressure, and $T_{\rm{phot}}(r)$ is given by Eq. (\ref{eq:Tphot}). For the surface density $\Sigma(r)$, we employ two different expressions: Eq. (\ref{eq:sigmaf}) above, and a power law used in semi-analytical calculations of a two-phase circumterrestrial disk [Machida and Abe, 2004 (MA04)]:

\begin{equation}\label{eq:sigmama1}
\Sigma_{\rm{MA}}(r)=\Sigma_{\rm{in}}\left(\frac{r}{r_{\rm{in}}}\right)^{-\delta},
\end{equation}

\begin{equation}\label{eq:sigmama2}
\Sigma_{\rm{in}}=\begin{cases}
  \frac{M_{\rm{disk}}}{2\pi r^{2}_{\rm{in}}}\left(\delta-2\right)\left[1-\left(\frac{r_{\rm{out}}}{r_{\rm{in}}}\right)^{-\delta+2}\right]^{-1}, & \text{if $\delta \neq 2 $}\\
  \frac{M_{\rm{disk}}}{2\pi r^{2}_{\rm{in}}}\left[\ln \left(\frac{r_{\rm{out}}}{r_{\rm{in}}}\right)\right]^{-1}, & \text{if $\delta=2$}

  \end{cases}
\end{equation}

\noindent in which $M_{\rm{disk}}$ is the total mass of the PLD, $\delta$ is a parameter that determines how compact the disk is, $r_{\rm{in}}$ and $r_{\rm{out}}$ are the inner and outer edges of the disk, respectively, and $\Sigma_{\rm{in}}$ is the surface density at the inner edge. In this case, we use $M_{\rm{disk}}=2M_{\rm{Moon}}$ (with $M_{\rm{Moon}}=7.35\times 10^{25}$ g), $r_{\rm{in}}=r_{\rm{E}}$,  $r_{\rm{out}}=10 r_{\rm{E}}$, and three values of $\delta$: 1, 2 and 3. Figure \ref{fig:alpha}a shows the resulting radial profiles of $\alpha$. The solid black curve uses Eq. (\ref{eq:sigmaf}) in Eq. (\ref{eq:alpha}) for $\alpha$, while the gray curves use Eqs. (\ref{eq:sigmama1}) and (\ref{eq:sigmama2}). The solid, dashed and dotted gray curves correspond to a disk compactness parameter $\delta$=1, 2 and 3, respectively. The MA04 values of the surface density produce $\alpha$ values that, for radii shorter than 7$r_{\rm{E}}$, are lower than the corresponding $\alpha$ values obtained from the TS88 expression of the surface density. In fact, within 1 Earth radius of the Earth's surface, the most compact PLDs ($\delta=$2, 3) may be too dense for dynamically significant turbulent activity, as indicated by the low values of $\alpha \sim 10^{-5}$ in that region. Nevertheless, between $r\sim2r_{\rm{E}}$ and $\sim 6r_{\rm{E}}$ the MA04 model of surface density gives $\alpha$ values that are more consistent with those obtained by Pahlevan and Stevenson (2007). The model of $\Sigma$ by TS88 yields $\alpha$ values between $\sim 10^{-4}$ and $\sim 3\times 10^{-3}$ along the radial extent of the disk.

We can also estimate upper bounds on $\alpha$ as a function of distance from the PLD mid plane, under the assumption that the disk gas is magnetized. Consider the ratio of gas pressure to magnetic pressure, or plasma beta, $\beta\equiv P/P_{\rm{mag}}$, where $P_{\rm{mag}}=B^{2}/8\pi$. In terms of the Alfv\'{e}n speed $v_{\rm{A}}$ (see Section \ref{sec:ion}),  the plasma beta can be written as

\begin{equation}\label{eq:beta}
\beta = \frac{2P}{\rho v^{2}_{\rm{A}}},
\end{equation}

\noindent and so

\begin{equation}
v_{\rm{A}}=\sqrt{\frac{2P}{\rho \beta}} \lesssim 2.43\times 10^{4} ~~ \rm{cm \,s^{-1}}.
\end{equation}

\noindent where the inequality was calculated above in (\ref{eq:vamax}). Therefore

\begin{equation}
\beta \gtrsim \frac{2\times(1.689\times 10^{-9}~\textrm{s$^{2}$ cm$^{-2}$}) P}{\rho}.
\end{equation}

\noindent But from Eqs. (\ref{eq:claus}) and (\ref{eq:rhoClaus}),

\begin{equation}
\frac{P}{\rho}=\frac{\ell T}{T_0}=(2.83\times 10^{6}~\textrm{erg g$^{-1}$ K$^{-1}$})T
\end{equation}

\noindent and hence 

\begin{equation}\label{eq:betaz}
   \beta \gtrsim 42 \, \left( \frac{ T_{\rm c} }{ 4200 \, {\rm K} } \right) \,
   \left[ 1 - \frac{1}{x_{\rm c}} \, \left( \frac{ z }{ H_0 } \right)^2 \right].
\end{equation}

\noindent In Eq. (\ref{eq:betaz}), the approximation (\ref{eq:Tz}) by Ward (2012) for the disk temperature was used. Finally, following Keith and Wardle (2014) and Hawley et al. (1995), we employ an empirical relation observed from numerical simulations of the MRI, in which the turbulence intensity and the saturated magnetic field are related by

\begin{equation}\label{eq:ab}
\alpha \approx \frac{1}{2\beta}
\end{equation}

\noindent to obtain

\begin{equation}\label{eq:az}
\alpha(z) \lesssim 1.2 \times 10^{-2} \, \left( \frac{ T_{\rm c} }{ 4218 \, {\rm K} } \right)^{-1} \,
 \left[ 1 - \frac{1}{x_{\rm c}} \left( \frac{z}{H_0} \right)^2 \right]^{-1}
\end{equation}

Figure \ref{fig:alpha}b shows vertical profiles of the upper limit given by Eq. (\ref{eq:az}) at $r=1.5 r_{\rm{E}}$ (black curve) and at $r=4r_{\rm{E}}$ (gray curve). While we are not calculating precise values of $\alpha$, this figure shows that its upper bounds, due to MRI turbulence, can provide efficient viscous transport of angular momentum in the PLD at different radii.

Taken together, the results in both panels of Fig. \ref{fig:alpha} point to a significant susceptibility of the PLD to turbulent flow regimes, whether they originate from the balance between viscous dissipation and photospheric losses (panel a)  or from the MRI (panel b). Since we are interested in the dynamical consequences of the latter mechanism, we carry out a numerical simulation to investigate its effects.

\section{MHD simulation}\label{sec:MHDsim}
In order to characterize the amount of mass transport that would be expected if the PLD were turbulent due to the MRI, we perform a numerical simulation of MRI turbulence in a simplified model of the protolunar disk. We employ the so-called shearing box approximation of an accretion disk (Hawley et al. 1995): a local patch of the disk, co-rotating with the flow, is represented as a rectangular Cartesian coordinate system, and is placed at a fiducial orbital radius $r_0$. The $x$, $y$ and $z$ directions in the box correspond to the radial, azimuthal and vertical directions in the disk, respectively. The differential rotation of the disk gas is replaced by a shear flow, which depends on $r_0$ through the angular frequency $\Omega$. 

Using this setup, we solve the equations of ideal MHD:

\begin{subequations}\label{eq:mhd}
\begin{align}
\frac{\partial \rho}{\partial t} + \nabla\cdot \left( \rho \boldsymbol{v}\right) &=  0\\
\frac{\partial \boldsymbol{v}}{\partial t} + \boldsymbol{v}\cdot \nabla \boldsymbol{v} &= -\frac{1}{\rho}\nabla \left( P + \frac{B^2}{8\pi}\right) + \frac{\left(\boldsymbol{B}\cdot \nabla\right) \boldsymbol{B}}{4\pi \rho} - 2\boldsymbol{\Omega}\times \boldsymbol{v} + 3\Omega^2 x \boldsymbol{\hat{x}}\\
\frac{\partial \boldsymbol{B}}{\partial t} = \nabla \times \left(\boldsymbol{v}\times \boldsymbol{B}\right)\\
\frac{ \partial \epsilon}{\partial t}=- P\nabla \cdot \boldsymbol{v}.
\end{align}
\end{subequations}

\noindent In these expressions, $\boldsymbol{v}$ is the gas velocity, $\boldsymbol{B}$ is the magnetic field, and $\epsilon$ is the internal energy density. The first equation expresses conservation of mass; the second equation, conservation of momentum, with the first term on the right-hand side corresponding to the force due to gas and magnetic pressure, the second term to the tension in the magnetic field lines, and the third and fourth terms the Coriolis force and the radial component of the Earth's gravity, respectively. The third equation describes the evolution of the magnetic field, while the fourth equation describes the evolution of the internal energy per unit volume. The system of equations needs to be closed by an equation of state (EOS). We follow Wada et al. (2006) and use an EOS that consists of two parts, an ideal gas term and a polytrope-like term,

\begin{equation}\label{eq:wkm}
P = \left(\gamma-1\right) \epsilon + \left(5-\gamma\right)\rho^5,
\end{equation}

\noindent where $\gamma$ is the ratio of specific heats. According to Wada et al. (2006), this EOS represents hot gaseous debris. Even though it may not be an altogether realistic representation of the geological materials present in the PLD, the simplicity of this EOS allows us to bring the effects of MRI turbulence into focus. 

Equations (\ref{eq:mhd}) and (\ref{eq:wkm}) are solved with the astrophysical MHD code ZEUS (Stone and Norman, 1992a,b), which employs a staggered mesh in which to evolve the relevant quantities. For the purpose of our simulation, the shearing box has dimensions $L_x=L_z=0.1r_{\rm{E}}$ and $L_y=0.6r_{\rm{E}}$, and the number of computational grid cells is $n_x\times n_y \times n_z=64\times 128\times 64$. The center of the box is placed at $r_0=4r_{\rm{E}}$. At this orbital radius, one orbit around the Earth ($2\pi/\Omega$) takes $\sim$11.3 hours. The ratio of gas to magnetic pressure is $\beta=10^3$. By virtue of Eq. (\ref{eq:beta}), this corresponds roughly to an Alfv\'{e}n speed $v_{\rm{A},z}\approx$ 1.5$\times10^5$ cm/s (using an initial gas pressure $P=1.99\times 10^8$ dyn/cm$^2$ and gas density $\rho=1.84\times 10^{-3}$ g/cm$^3$), which follows relation (\ref{eq:vamax}).

A very important parameter in shearing box simulations is the initial configuration of the magnetic field threading the box. In this simulation, the initial field is vertical and sinusoidally-varying along the radial $x$ direction. This is known as a zero-net magnetic flux case. MHD turbulence in this instance is self-sustained through a local magnetic dynamo, and is generally known to be weaker than turbulence produced by an initially uniform, non-zero-net magnetic flux (see Turner et al. 2014a for a review). In this sense, our simulation represents a "worst-case scenario", if we assume that the whole disk is fully ionized: it will always give a smaller value of $\alpha$ than if we had used a uniform magnetic field. Although the initial magnetic field topology has a direct bearing on the subsequent evolution of the MHD flow, it is beyond the scope of this work to perform a detailed analysis of the different magnetic configurations, or to carry out an extensive parameter survey. These will be deferred to future studies.

\begin{figure}[h!]
\begin{center}
\includegraphics[width=0.9\textwidth]{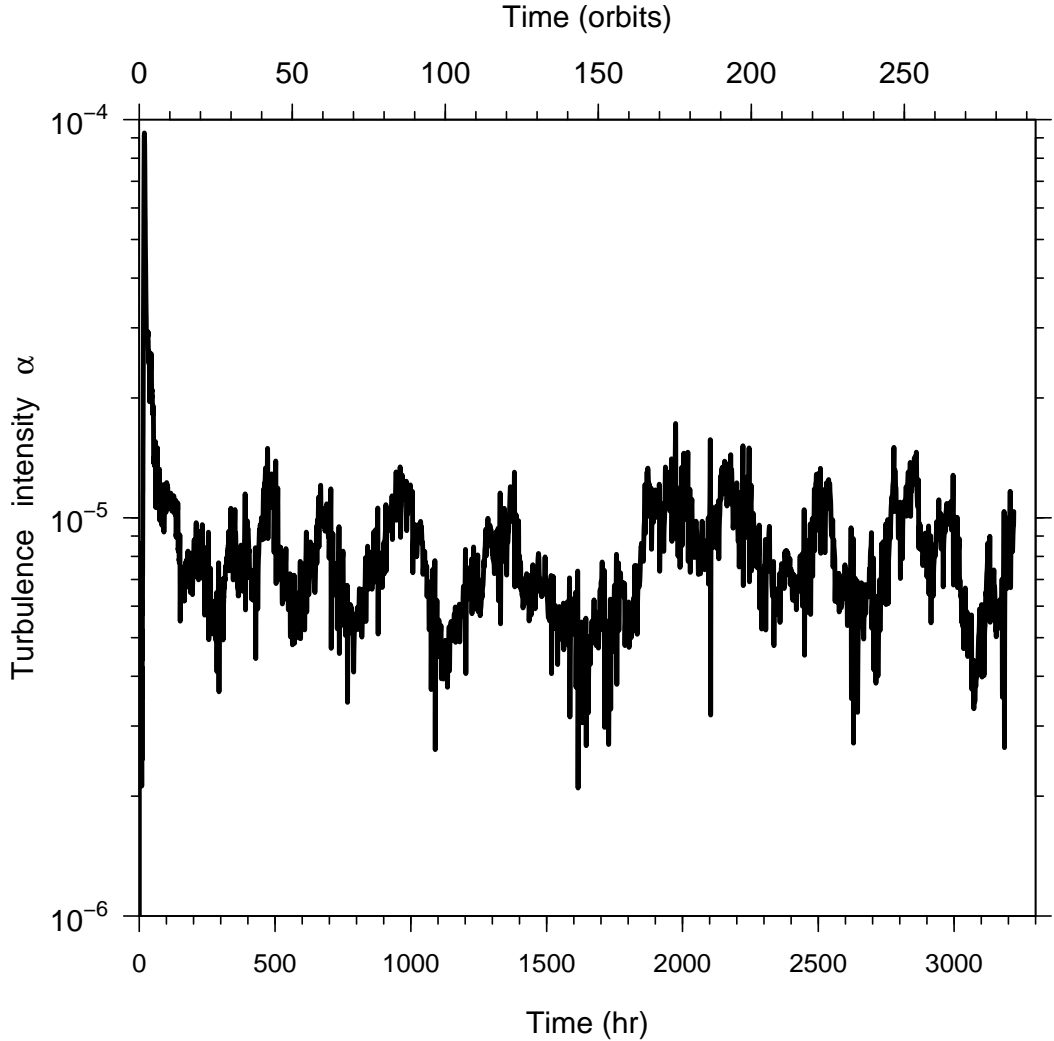}
\end{center}
\caption{\small{Time evolution of the turbulent intensity $\alpha$. The $\alpha$ parameter is obtained by volume-averaging the magnetic and the hydrodynamic stresses in the shearing box, adding them together and normalizing the result by the initial gas pressure.  }}    \label{fig:hst}
\end{figure}

Figure \ref{fig:hst} shows the time history of the turbulent $\alpha$ parameter from the simulation. To calculate $\alpha$, the volume-averaged magnetic and hydrodynamic stresses are added together,  and the sum is divided by the initial gas pressure. The rise of $\alpha$ early on is associated to a "channel" flow, which consists of two oppositely directed streams and is an exact nonlinear solution of the shearing box equations. In a channel flow, the magnetic perturbations and the velocity field experience a high degree of spatial correlation, resulting in increased stresses (Balbus \& Hawley 1998). This mode begins to oscillate vertically after about 2 orbits, and breaks up into a turbulent state. Thereafter, $\alpha$ oscillates around $\sim 7\times 10^{-6}$ for the duration of the simulation.

Despite the relatively low values of turbulent activity observed in this particular simulation, diffusion of isotopic species due to MRI turbulence may still be achieved in short enough times to homogenize portions of the PLD. To see this, we introduce a passive contaminant in the turbulent flow and track its radial motion across the box. The procedure is similar to that followed by Carballido et al. (2005) in a more general accretion disk context. 

\begin{figure}[h!]
\begin{center}
\includegraphics[width=0.99\textwidth]{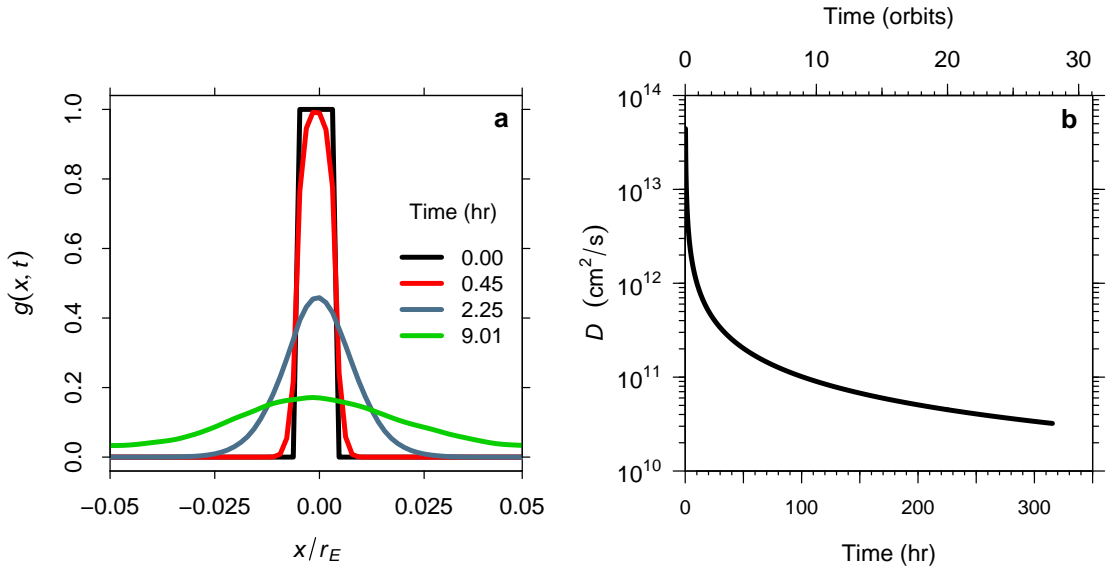}
\end{center}
\caption{\small{\textbf{(a)} Azimuthally- and vertically-averaged radial profiles of the contaminant fraction $g(x,t)$, shown at different times between 0 and 9 hours after being introduced in the turbulent flow; \textbf{(b)} Diffusion coefficient $D$ of the contaminant obtained by fitting the analytical solution of a diffusion equation for $g$ to numerical profiles of the form shown in (a). }\label{fig:fprof}}
\end{figure}

The contaminant is modeled as a fraction $g(x,t)$ of the gas density, and is evolved by solving a continuity equation for $g\rho$. The initial condition on $g$ is a constant value of 1 within 0.005$r_{\rm{E}}$ of the $x=0$ plane of the shearing box. The contaminant is initialized in the turbulent flow, with its original distribution, every 3 orbits (34 hours) after the first 10 orbits of shearing box evolution. After being introduced, the contaminant is allowed to spread for 30 orbits. This is done 10 times in order to improve our measurements, since we anticipate large statistical fluctuations in the evolution of $g$ due to the turbulence. To obtain radial profiles of the contaminant fraction, we average it over the azimuthal ($y$) and vertical ($z$) directions. Each radial profile is further averaged over the number of times that we re-introduced the contaminant. The results are shown in Fig. \ref{fig:fprof}a.

To obtain a more quantitative idea of the transport rate, we model the turbulent mixing of the contaminant as a diffusion process with diffusion coefficient $D$, and we solve a diffusion equation for $g$,

\begin{equation}\label{eq:diffusion}
\frac{\partial g}{\partial t}=D\frac{\partial^{2} g}{\partial x^2}
\end{equation}

\noindent subject to the initial condition $g(x,0)=1$ for $|x| \leq 0.005r_{\rm{E}}$, and 0 for $|x| >0.005r_{\rm{E}}$. 

The solution is then fitted to the profiles measured from the simulation, with the diffusion coefficient used as the fitting parameter. The fitting procedure is done independently for each time of the contaminant evolution. The values of $D$ calculated in this manner are plotted in Fig. \ref{fig:fprof}b. According to our simplified model, the diffusivity of the contaminant decreases considerably during the first orbital period, and more slowly thereafter. The reason for this decline may be due to an increasingly large Peclet number $Pe\equiv ul/D$, where $u$ is the characteristic flow velocity and $l$ is a length scale of the order of the box size, as the contaminant is spread over long length scales by large turbulent eddies. We also note, however, that Eq. (\ref{eq:diffusion}) may not describe the mixing process completely (e.g., Hubbard and Brandenburg, 2009). We therefore use a range of values of $D$ to estimate the mixing time scale. With $D \sim 10^{11}$ cm$^2$ s$^{-1}$, the diffusion time of a passive scalar across a radial distance 0.1$r_{\rm{E}}$ is (0.1$r_{\rm{E}})^2$/($10^{11}$ cm$^2$ s$^{-1})\approx$ 11.3 hr, and across 10$r_{\rm{E}}$ it is $\sim 13$ yr. If we use $D\sim 10^{10}$ cm$^2$ s$^{-1}$, we obtain 113 hr and 129 yr, respectively. These figures are below the depletion time of the disk vapor, $\sim250$ years (TS88, Ward 2012). Nevertheless, once the diffusion coefficient reaches $\sim 5\times10^{9}$ cm$^2$ s$^{-1}$, the mixing time over radial distances of the order of $\sim10r_{\rm{E}}$ becomes longer than the disk life time.

\section{Discussion}\label{sec:discussion}
\subsection{Recapitulation}
In light of the very high temperatures that were reached in the protolunar disk according to previous estimates, we have combined analytical models of the structure of the protolunar disk by TS88 and by Ward (2012) with results from chemical calculations by Visscher and Fegley (2013), with the aim of estimating ionization fractions in the gaseous component of the PLD. We found that for some chemical species (prominently Na and K, but also SiO, O$_2$ and Zn) those fractions are above the minimum one required for the onset of the magneto-rotational instability, a magnetohydrodynamic mechanism that transports angular momentum outwards in a differentially-rotating system, such as the PLD, through turbulence. If differential rotation and a sufficient level of ionization are present, a relatively weak magnetic field is all that is needed to trigger this instability. Remarkably, the MRI has not been considered in the PLD before. 

We have determined, for the first time, that MRI turbulence may have produced very efficient mixing of isotopic and chemical species within the PLD (note that this process is independent of the possible mixing between the disk and the Earth). We base this claim on our estimates of the strength of the turbulence in the disk, parameterized by the dimensionless number $\alpha$ of accretion disk theory. Our estimates of $\alpha$ yield high values of several times $10^{-2}$, particularly at high disk altitudes. In contrast, a numerical MHD simulation of a local patch of the PLD gives $\alpha \sim 10^{-5}$ at the most. We emphasize that this low numerical value is a result of the initial magnetic field geometry used in our simulation: our so-called zero net magnetic flux  configuration produces weaker turbulence than non-zero net magnetic flux calculations, for reasons that are still being investigated (Turner et al. 2014a). Shearing box size and resolution can also play a role (Sec. \ref{sec:caveats}). Nevertheless, we have shown that a turbulence level this low is still sufficient to effectively mix a passive scalar (a dynamical proxy for isotopic species) over radial length scales of several Earth radii within the lifetime of the gaseous PLD.

We can make a rough estimate of the magnetic field strength that would have driven MRI turbulence in the PLD. Using Eq. (\ref{eq:ab}) and the definition of the plasma beta, $8\pi P/B^{2}$, we write

\begin{equation}\label{eq:magfield}
B\approx \sqrt{16\pi \alpha P}
\end{equation}

\noindent With a value of $T_{\rm{phot}}\approx 2000$ K, we have $P= P_0\mathrm{exp}(-T_{0}/T_{\rm{phot}}) \approx 28$ dyn cm$^{-2}$. From our simulation, we obtained $\alpha \sim 10^{-5}$ (Fig. \ref{fig:hst}), and from Eq. (\ref{eq:az}) an approximate upper value of $\alpha \sim 10^{-2}$ was estimated. We therefore have $B\approx 0.12$--4 G. For comparison, field intensities obtained from 3.45 Ga quartz phenocrysts yield $B\approx 0.18$--0.28 G (Tarduno et al. 2014).

\subsection{Implications for abundances of volatiles }\label{sec:implications}
As already stated, our calculations do not explicitly address mixing between the PLD and the Earth. It is possible that a gaseous envelope could surround both the Earth and the PLD, allowing thorough mixing and isotopic homogenization that may explain the surprisingly uniform isotopic compositions in the Earth and Moon (Fig. 1). On the other hand, physical arguments suggest the mixing between the Earth and PLD may be self-limiting (Desch and Taylor, 2013), and mixing between the Earth's surface and the deep mantle is not predicted (Nakajima and Stevenson, 2014). The extended period of turbulent mixing in the hot disk would likely have been accompanied by continuous loss of volatile elements from the photosphere. We have not modeled this aspect of the problem yet, but abundances of alkali elements suggest loss of about 80\% of terrestrial values and loss of about 99\% of highly volatile elements, including Zn. If driven by Rayleigh fractionation where elements are essentially boiled off the top of the photosphere, such high losses should have led to substantial isotope fractionation. For example, Paniello et al. (2012) show that Rayleigh fractionation of Zn isotopes, if Zn is lost as ZnCl$_2$, would lead to the observed $\delta$\ce{^{66}Zn} by about 10\% loss. Using the same procedures, we calculate that if Zn is lost as elemental Zn, which has a lower fractionation factor, only 5\% loss would be required. These losses are substantially below the 99\% depletion in Zn in the Moon compared to Earth. Thus, the primary loss mechanism must be one that does not lead to significant mass fractionation, such as hydrodynamic escape (Desch and Taylor, 2011). The immeasurable isotopic fraction of K isotopes is consistent with a loss mechanism not involving Rayleigh fractionation. The small difference in Zn isotopes between Earth and Moon might have been caused by short episodes of Rayleigh fractionation, but one cannot rule out isotopic fractionation after lunar formation. 

Both concentrations and isotopic compositions of volatile elements generally show that the Moon is depleted in volatile elements and that these elements are enriched in their heavy isotope compared to Earth. There are exceptions, however. For example, the region of the lunar interior that gave rise to orange glass pyroclastic deposit is not depleted in volatiles and has light Zn isotopic composition ($\delta$\ce{^{66}Zn} of $<-3.5$; Kato et al., 2015). The low $\delta$\ce{^{66}Zn} might reflect Rayleigh loss during eruption, but the high concentrations of volatile elements indicates formation of portions of the Moon from materials that had not yet lost volatiles from the PLD. Salmon and Canup (2012) suggest that moonlets could form rapidly (< 50 years) in the outer reaches of the PLD, followed by others that were more thoroughly processed (hence degassed) in the disk. Recently, Canup et al. (2015) showed through a combination of thermal, chemical and dynamical models that volatile depletion in the Moon may be explained by accretion of volatile-rich melt onto the Earth, instead of on the growing Moon. 

In summary, these considerations suggest that the PLD was efficient at mixing throughout its volume, both radially and vertically. Equilibration with the Earth is likely. Loss of volatiles could have occurred by gas loss from the surface of the PLD by a mechanism that did not significantly alter the isotopic compositions of volatiles.

\subsection{Caveats and future work}\label{sec:caveats}
The calculations reported here represent a preliminary investigation of the MHD structure of the protolunar disk. Several important pieces of physics remain missing. Among these, the coupling of the MRI to the energy balance between viscous dissipation and radiation is crucial. Note that Eqs. (\ref{eq:mhd}) do not include these effects, which TS88 and other authors have shown dominate the evolution of the PLD. 
 
A major assumption in this work is that the disk was completely gaseous. However, most analytical and numerical models consider a two-phase structure in which a disk of hot, liquid magma was surrounded by vapor. Furthermore, a thermodynamic treatment shows that magma droplets may have condensed out of the gas atmosphere and replenished the liquid layer (Ward 2012). A more complete magnetohydrodynamic study would need to take this structure into account, since the magma sub-disk may have become decoupled from the MRI-active gas disk, and therefore it may have mixed less. This is an important point, because the Moon likely formed from the dense magma sub-disk by gravitational instability. Moreover, the dynamical decoupling between liquid and vapor could have given rise to effects such as a Kelvin-Helmholtz instability, which may have nonetheless mixed the two components vertically and perhaps delayed the formation of the Moon. An analogous effect is predicted for protoplanetary disks (Cuzzi et al. 1993), in which a disk of solid particles orbiting at Keplerian speed interact with the surrounding disk gas, which in turn orbits at a sub-Keplerian speed due to an existing radial pressure gradient. The Kelvin-Helmholtz turbulence that ensues prevents the gravitational collapse of the disk of solids, and planetesimal formation is quenched. 

It may be natural to inquire whether the existence of droplets suspended in the gas phase could have played a role in regulating the ionization of the PLD. Could adsorption of electrons and ions onto magma droplets perhaps reduce the ionization fraction below the critical value for MRI development? The answer may be negative: thermionic emission of electrons and ions from magma droplets is likely to \textit{increase} the ionization fraction, especially at temperatures $\gg 2000$ K (Desch and Turner, 2015), so recombination reactions on droplet surfaces probably do not dominate at such high temperatures. Further investigations should take into account thermionic emission from droplets.

When considering MHD processes in systems such as protoplanetary and circumplanetary disks, it becomes necessary to quantify non-ideal effects such as ambipolar diffusion. However, it is unlikely that the rate of magnetic diffusion including ambipolar diffusion would be significantly greater than that due to Ohmic dissipation alone in the PLD (as assumed in Section \ref{sec:ion}). Ambipolar diffusion dominates
 over Ohmic dissipation only when $\omega_{\rm i} / \nu_{\rm i} \gg 1$, where $\omega_{\rm i} = e B / m_{\rm i}$ is the gyrofrequency of
 ions of mass $m_{\rm i}$ around the magnetic field $B$, and $\nu_{\rm i} = n_{\rm n} \langle\sigma v\rangle_{\rm i,n}$ is the collision frequency of ions with neutral gas molecules of number density $n_{\rm n}$. The relative speed between ions and neutrals is $v$ and their collision cross section is $\sigma$.  Assuming $\langle \sigma v\rangle_{\rm i,n} = 3.5 \times 10^{-13} \, {\rm cm}^{3} \, {\rm s}^{-1}$ (Draine et al. 1983),
 $n_{\rm n} \sim 10^{20} \, {\rm cm}^{-3}$,  and $m_{\rm i} = 23 m_{\rm p}$, magnetic field strengths $B$ would need to exceed
 an implausible $\sim 10^5 ~{\rm G}$ for ambipolar diffusion to dominate.

The use of the equation of state (\ref{eq:wkm}) by Wada et al. (2006) was called into question by Nakajima and Stevenson (2014), who argue that such a polytrope-type EOS will fail to model phase changes of mantle materials, particularly in the non-equilibrium processes associated to the Moon-forming giant impact. The shocks that were observed in the PLD by Wada et al. (2006), as a result of large density contrasts, led to a very rapid rate of angular momentum transport and disk evolution, and consequently Moon formation by gravitational instability under those circumstances was deemed to be unlikely. The numerical MHD model studied here does not develop substantial density differences, and represents an approximate dynamical steady state for which the EOS (\ref{eq:wkm}) is adequate. Nevertheless, a more sophisticated EOS would allow to capture relevant properties of PLD materials. For example, Canup et al. (2013) studied potential lunar-forming impacts that generated a variety of circumterrestrial disks, employing both a smoothed particle hydrodynamics method and a grid-based code. The equation of state that they used, called ANEOS, is a semi-analytic algorithm in which thermodynamic quantities are derived from the Helmholtz free energy, with temperature and density as independent variables. A revised version, called M-ANEOS, includes the formation of silicate vapor composed of diatomic molecules, such as MgO or SiO. Although ANEOS is primarily used for impact calculations, it may allow for a better treatment of gas thermodynamics in an MHD model of the PLD, and it would provide accurate determination of two-phase regions (liquid and vapor), if a good understanding of the development of the MRI in a two-phase system is achieved beforehand.

Among the parameters that determine the level of MRI turbulence in shearing box models of accretion disks, the size of the box plays an important role. If the initial vertical field produces a non-zero net magnetic flux through the box, the MHD stresses experience episodic activity in the form of channel flows (Section \ref{sec:MHDsim}) if the box aspect ratio between the radial and vertical box dimensions is close to 1 (Bodo et al. 2008). As the aspect ratio increases, the channel flows die out and the turbulent intensity $\alpha$ converges towards smaller values. Numerical resolution can also modify the amount of turbulent transport: in the absence of a net magnetic flux, an increase in resolution can lead to a decrease in $\alpha$ (Bodo et al., 2011). Moreover, modeling the vertical variation of the density of the disk material (i.e., making the box several scale heights tall and including the vertical component of Earth's gravity), which is necessary to understand the behavior of $\alpha$ with height, introduces additional constraints on the grid resolution, since high Alfv\'{e}n speeds in the uppermost disk regions require short time steps. Clearly, numerical studies of turbulent mixing in the PLD using the shearing box formalism may still yield an extensive range of possible results. 

The diffusive model of Pahlevan and Stevenson (2007) indicates that mixing in the Moon-forming disk can equilibrate its isotopic composition in $\sim100$ years, if $\alpha\approx 3\times10^{-4}$. In fact, this diffusion time may be somewhat uncertain, given that the authors assumed equality between the turbulent diffusion coefficient $D$ and the effective kinematic viscosity due to turbulence, $\nu_{\rm{t}}$, in Eq. ({\ref{eq:alphat}). There is a subtle difference between diffusivity and viscosity in an accretion disk: while diffusive mixing (as measured by a diffusion coefficient $D$) tends to produce a uniform concentration of a passive contaminant, the viscosity that arises due to turbulence does not tend to produce a uniform distribution of angular momentum. This difference can be quantified by the Schmidt number, $\textrm{Sc}\equiv \nu_{\rm{t}}/D$. Values of $\rm{Sc}$ that have been measured using MHD simulations range from $\sim 0.1$ to $\sim 10$. So, while a value of $\rm{Sc}\sim 1$ such as assumed by Pahlevan and Stevenson (2007) cannot be ruled out, the dynamics of turbulence in the PLD could produce faster or slower homogenization than $\sim100$ years, depending on whether transport of mass or transport of angular momentum dominates, respectively.

\section{Conclusions}\label{sec:conc}
We performed the first study of possible magnetohydrodynamic effects in the protolunar disk, using a model in which we decouple the liquid phase from the gaseous phase, and focus on the latter. Our findings are the following:
\begin{enumerate}
\item The most abundant chemical species in the protolunar disk according to published data, Na and SiO, are among the most significant contributors to the ionization fraction of the gas component of the disk. Although K has a relatively low abundance, its low ionization potential also makes it a prime contributor to the ionization budget.

\item The values of the ionization fraction above the minimum necessary for the onset of magneto-rotational instability occur in the disk photosphere (at all disk radii), and all along the vertical extent of the disk.

\item The strength of the turbulence in the protolunar disk due to the magneto-rotational instability has values $\alpha\lesssim 10^{-2}$, and increases with disk height more sharply at large orbital radii.

\item A numerical magnetohydrodynamic simulation of a small patch of the disk, using a polytrope-like equation of state for the gas and an initial vertical magnetic field with vanishing flux, gives a time-averaged value of $\alpha \approx 7\times10^{-6}$. Despite this low value, turbulence would be able to mix a passive tracer (used as a proxy for isotopic or chemical species) across a radial distance of 10 Earth radii within the lifetime of $\sim$ 250 yr of the gas disk, provided the turbulent diffusion coefficient does not decrease too much below $\sim 10^{10}$ cm$^2$ s$^{-1}$.

\item Future studies should take into account: \textit{i)} the two-phase structure of the PLD, using a more realistic equation of state (such as ANEOS);  \textit{ii)} the energy balance between radiation and viscous dissipation; and, if shearing box calculations are performed, \textit{iii)} the role of various numerical parameters like initial magnetic field geometry, vertical stratification, resolution, and plasma beta.

\end{enumerate}

\subsection*{Acknowledgements}
The input received by D. Stevenson and an anonymous referee helped improve the descriptions of the microphysical and geochemical aspects of this study. This work benefited greatly from stimulating conversations with Y. Nakamura, C. Frohlich, and K. Soderlund. GJT thanks the NASA Solar System Exploration Research Virtual Institute for support under Cooperative Agreement Notice NNA14AB07A (D. Kring, PI).

\end{document}